\documentstyle[12pt]{article}
\RequirePackage{graphicx}
 \RequirePackage{times}
\RequirePackage{mathptm}
\begin{document}

\sloppy
\title{What is wrong with SLASH\thanks{SLASH: acronym for Second
 Laser--Amplified Superluminal Hookup}$\;$  ?}
\author{Karl Svozil\\
 {\small Institut f\"ur Theoretische Physik, University of Technology Vienna }     \\
  {\small Wiedner Hauptstra\ss e 8-10/136,}
  {\small A-1040 Vienna, Austria   }            \\
  {\small e-mail: svozil@tuwien.ac.at}}
\maketitle

\date{ }
\maketitle

\begin{abstract}
 In an experiment featuring
 nonlinear optics, delayed choice and EPR-type correlations,
the possibility of faster--than--light communication appears not totally implausible.
 Attempts are put forward and discussed to refute this claim.
\end{abstract}

 Quantum theory and special relativity theory are the cornerstones
 of today's physical perception.
 Obviously their intrinsic consistency as well as the consistency of a
 combined theory is of greatest relevance.
 Thus one of the most discomforting results of 20th century mathematics and physics
 is the conclusion that {\em with respect to consistency of theories,
 in general no affirmative answer can be given}---
 It must be clearly spelled out that the mere introduction of
 ``manifestly covariant'' entities, such as Lorentz invariant
 spinors and tensors
 in  quantum electrodynamics, is an
 insufficient guarantee of consistency.
 Any attempt to ``prove'' consistency in nontrivial theory contexts,
such as for instance Shirokov's
 remarkable investigation \cite{shirokov},
 must inevitably be too specific or even misleading.

Although eminent physicists and philosophers of science have
conjectured the ``peaceful coexistence doctrine'' \cite{shimony2} of quantum and relativity theories,
 it cannot be excluded that inconsistencies may be ``lurking behind the corner.''
 With much luck and a little intuitive insight every layman may produce them.
And although just like in mathematics, physicists are assured
every laboratory day since almost a century, that there is no such problem,
 they are in the somewhat discomforting position.
A single experiment,
 which is explicitly spelling out the inconsistency of quantum
 mechanics with special relativity theory, may once and for all
 discredit the peaceful consistency doctrine.
 Therefore,
 it seems not totally unjustified to search for relevant counterexamples, at least for
 the reason to obtain new insights by disproving them.

      Earlier attempts \cite{herbert} to utilize nonlinear optical
 devices for signal amplification (``photon cloning'') to construct a
 FLASH\footnote{ FLASH: acronym for First Laser--Amplified Superluminal
 Hookup, see ref. \cite{herbert}.} failed, heuristically speaking,
 because of the
 impossibility of noiseless amplification \cite{glauber} (see also
 \cite{wo-zu,mil-hard,mandel:83,bussey})
and the reversivility (i.e., one-to-oneness) of the unitary
quantum state evolution.
In this Communication a proposal for another
 arrangement will be presented which at first glance seems to be able
 to utilize delayed choice on the basis of an EPR--type configuration
 for faster--than light signalling.

      The proposed device consists of a photon pair source, polarizers
 and detectors.  Let $\theta$ be the relative angle of the polarizers.
 If the photon pairs emitted by the source are in a total angular
 momentum 0 and total parity $+1$ state,
 the normalized detection rate relative to the counting rate  without polarizers
 is given by $\cos^2 (\theta )/2$ \cite{clauser}.

      Consider now the same arrangement with the following change.
 Instead of the polarizer and the detector a nonlinear device is
 installed on either one of the photon paths. The nonlinear device
 could be thought of as a {\it box containing a great number of
 photons, all of them
 in the same state}. A particular realization would be a
 polarized laser. This device has the
 property of amplifying the scattering amplitude of the corresponding
 photon into a specific polarization state.

      In a  model  it is assumed that there are
 $n$
 quanta with polarization ``$\updownarrow$'' present in the nonlinear
 device.  Let $\vert \updownarrow i\rangle $ and $\vert \leftrightarrow
 i\rangle $ denote the state of the $i$'th photon for $n=0$ with
 polarization ``$\updownarrow $'' and ``$\leftrightarrow $'',
 respectively.
 The corresponding states with a population of $n$ quanta are denoted
 by $\vert
 \cdot , n\rangle $. The normalized conditional amplitude that a
 quantum will be
 scattered into a state when there are already $n$ quanta present, is
 \cite{feynman-III} $\langle n+1\vert n\rangle =(n+1)^{1/2}$.  Then the
 emitted pair wave function can be written as ($N$ is a normalization
 constant)
 \begin{eqnarray}
 \vert \Psi ,n\rangle &=&{1\over N} \left( \vert \updownarrow
 1,n\rangle \vert \updownarrow 2\rangle +\vert \leftrightarrow 1\rangle
 \vert \leftrightarrow 2\rangle \right) =\nonumber \\
&=&{1\over N} \left(
   \langle n+1\vert n\rangle
 \vert \updownarrow 1 \rangle   \vert
 \updownarrow 2    \rangle
 +\vert \leftrightarrow 1\rangle \vert \leftrightarrow
 2\rangle \right) =\nonumber \\
&=&{1\over (n+2)^{1/2}} \left[ (n+1)^{1/2} \vert
 \updownarrow 1\rangle \vert \updownarrow 2\rangle + \vert
 \leftrightarrow 1\rangle \vert \leftrightarrow 2\rangle
 \right] \qquad .
 \end{eqnarray}

      Consider the following limits: {\it (i)} $n=0$ corresponds to the
 situation with no amplifying photons in the nonlinear device.  This
 results in the usual pair wave ansatz yielding the
 above mentioned relative counting rates of $\cos^2(\theta )/2$;
 {\it (ii)}
 $n\longrightarrow
 \infty$ corresponds to an ideal amplifier.  Due to the statistics of a
 ``very large number'' of Bose quanta, $\vert \Psi
 ,n\longrightarrow \infty \rangle
 \longrightarrow \vert \updownarrow 1\rangle \vert \updownarrow
 2\rangle $.

\begin{figure}
\begin{center}
\unitlength 1mm
\linethickness{0.4pt}
\begin{picture}(126.67,47.33)
\put(50.00,20.00){\circle{10.00}}
\put(50.00,20.00){\makebox(0,0)[cc]{S}}
\put(55.00,20.00){\vector(1,0){20.00}}
\put(75.00,20.00){\vector(1,1){15.00}}
\put(75.00,20.00){\vector(1,-1){15.00}}
\put(110.00,5.00){\oval(30.00,10.00)[]}
\put(121.0,5.00){\oval(11.67,10.00)[l]}
\put(105.33,5.00){\makebox(0,0)[cc]{NL}}
\put(95.00,27.33){\line(0,1){14.67}}
\put(95.00,42.00){\line(1,1){5.00}}
\put(100.00,47.00){\line(0,-1){15.00}}
\put(100.00,32.00){\line(-1,-1){5.00}}
\put(97.33,37.33){\makebox(0,0)[cc]{P}}
\put(110.00,35.00){\oval(10.00,10.00)[r]}
\put(110.00,30.00){\line(0,1){10.00}}
\put(112.00,35.00){\makebox(0,0)[cc]{D}}
\put(91.00,35.00){\vector(1,0){18.67}}
\put(21.67,12.33){\line(0,1){14.67}}
\put(21.67,27.00){\line(1,1){5.00}}
\put(26.67,32.00){\line(0,-1){15.00}}
\put(26.67,17.00){\line(-1,-1){5.00}}
\put(24.00,22.33){\makebox(0,0)[cc]{P}}
\put(6.67,20.00){\oval(10.00,10.00)[l]}
\put(6.67,15.00){\line(0,1){10.00}}
\put(4.67,20.00){\makebox(0,0)[cc]{D}}
\put(45.00,20.00){\vector(-1,0){38.33}}
\put(75.00,22.67){\vector(-1,-4){0.2}}
\bezier{96}(73.00,33.00)(78.67,38.67)(75.00,22.67)
\bezier{116}(73.00,33.00)(67.33,24.33)(70.00,42.33)
\put(69.33,47.33){\makebox(0,0)[cc]{choice}}
\end{picture}
\end{center}
\caption{\label{1989-slash-f1} EPR--type configuration: entangled photons are emitted from the  source S.
In the right beam path, there is a choice between a polarizer (P) and a subsequent detector (D) or
 a nonlinear device (NL) for amplifying the amplitude of one polarization
 direction.}
\end{figure}
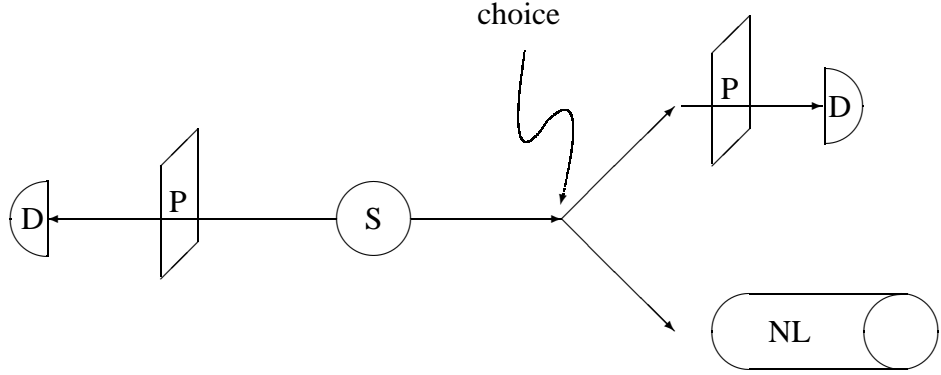

      Assume now that the nonlinear device can be arbitrarily
 positioned in or out of the beam.  Assume further that it
 has
 absolute efficiency, i.e. {\sl all} photons on one path are prepared
 (``materialize'') in a specific direction of polarization.  Let the
 angle between this direction of polarization and the polarizer at the
 path of the other photon be zero (or $\pi /2$) --- this guarantees
 maximal correlation.  Such an arrangement is drawn in Fig. \ref{1989-slash-f1}.

      It looks as if information could be transferred instantaneously
 from a spacetime region where the nonlinear device is positioned to
 the region of the polarizer at the end of the path of the other
 photon.  This can be done by simply hooking the device in and out of
 the photon path on the one side, corresponding to different relative
 photon detection rates of 1 (or 0 for $\theta =\pi /2$) and 1/2 on the
 polarizer side (with respect to the detection rate without polarizer).

      The above arrangement is different from FLASH \cite{herbert} since
 it requires only signal amplification in one photon direction and no
 ``generic photon cloning,'' producing
 noise and destroying coherence between the photon pair.
 In contradistinction, the present argument is based on
 the fact that the impossibility to control undecidable (``random'')
 single quantum events in a sense ``saves'' quantum theory from
 acausality.  It is an attempt to actively gain control over such
 single events by {\em stimulated emission}, thereby producing {\sl parameter
 dependence} of probability amplitudes at spatially separated points.

This is of course violating the fundamental but unproven conjecture
 that quantum theory is consistent with relativity theory,
to which the author adheres.
In what follows, four attempts will
 shortly be discussed to argue against faster--than--light
 communication with SLASH.

      {\it (i)} One might argue that the ansatz (1)---(3) is rather {\it ad hoc}
and  not based
 upon any trustworthy (quantum amplification) model, such as Glauber's
 amplifier model.  In this view, one incorrectly does not take into account
 the correct signal--to--noise ratio.  For
 instance, one might put forward that the nonlinear device would
 produce additional noise in the form of photons directed towards the
 other beam, thereby scrambling the correlated pair signal.  This could
 be circumvented by the insertion of a filter between the source and
 the nonlinear device which is transparent for the photons from the
 source but reflects photons coming from the nonlinear device.
 However, whether such a semitransparent filter can be made noiseless
 is an open question.

 {\it (ii)}
 One might perceive the production of the EPR correlated photon pair at
 the source as somehow ``actual'', in the sense that the
 direction of polarization of each correlated photon pair at the source
 is independent of what happens with either one or both of the beams
 ``afterwards''. This essentially boils down to local realism.
 Although local realism will not be critically discussed here, it
 should be mentioned that
 the configuration is similar to a delayed choice type experiment,
 where this argument fails.

 {\it (iii)} One modification of the above argument is that since
 (1)---(3) are essentially valid for processes in which quanta are
 produced and scattered into $n$--quanta states only if these
 subsystems
 are ``very close by'', i.e., not spatially separated.  The statistical
 properties of manybody processes decrease with the distance of two
 subsystems and effectively vanish for all practical purposes.
The exact form of such a signal attenuation remains unknown and may be the subject of further
studies.

 {\it (iv)}
 A further attempt to disprove SLASH has been put forward
 \cite{herbert-priv}: assume a modification of the previous setup
 by
 insertion of an anomalous refractor such as a calcite crystal into the
 beam pass of the (optional) nonlinear device. The calcite would
 split the beam, directing photons of one polarization direction to a
 properly adjusted
 nonlinear device which could be effectively perceived as a
 polarization analyzer, and photons of the
 other beam (with perpendicular polarization direction) to a detector.
 As a consequence of
 the perception of the nonlinear device as a passive analyzer,
 the SLASH requirement as an active element of the setup would then be
 discredited.

\begin{figure}
\begin{center}
\unitlength 1mm
\linethickness{0.4pt}
\begin{picture}(90.00,51.33)
\put(5.00,15.00){\circle{10.00}}
\put(5.00,15.00){\makebox(0,0)[cc]{S}}
\put(15.00,25.00){\line(1,-2){10.00}}
\put(70.00,40.00){\oval(10.00,10.00)[r]}
\put(70.00,35.00){\line(0,1){10.00}}
\put(80.00,25.00){\oval(20.00,10.00)[]}
\put(85.00,25.00){\oval(10.00,10.00)[l]}
\put(75.00,25.00){\makebox(0,0)[cc]{NL}}
\put(72.33,40.00){\makebox(0,0)[cc]{D}}
\put(70.00,5.00){\oval(10.00,10.00)[r]}
\put(70.00,0.00){\line(0,1){10.00}}
\put(72.33,5.00){\makebox(0,0)[cc]{D}}
\put(10.00,15.00){\vector(1,0){10.00}}
\put(20.00,15.00){\vector(3,1){40.00}}
\put(60.00,28.33){\vector(1,1){7.67}}
\put(60.33,28.33){\vector(2,-1){7.33}}
\put(20.00,15.00){\vector(4,-1){47.67}}
\put(60.00,31.67){\vector(0,-1){0.2}}
\bezier{88}(57.00,39.00)(61.67,45.33)(60.00,31.67)
\bezier{84}(57.00,39.00)(53.00,34.00)(54.33,49.00)
\put(54.00,51.33){\makebox(0,0)[cc]{choice}}
\put(12.67,29.33){\makebox(0,0)[cc]{calcite}}
\end{picture}
\end{center}
\caption{\label{1989-slash-f2} Delayed choice configuration featuring nonlinear photon gaintube.}
\end{figure}
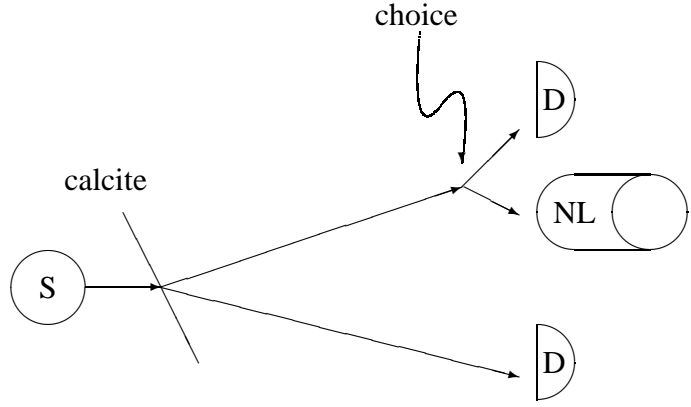

 To stress this point, a simplified version of the
 experimental setup is suggested (see Fig. \ref{1989-slash-f2}), in which the EPR--source
 is substituted by an unpolarized photon source. This source is
 directed towards an anomalous refractor.
 One outcoming beam is then directed towards a photon detector, whereas
 the other beam is {\it optionally} directed
 towards a nonlinear device (whose polarization axis is
 parallel
 to the polarization axis of the second beam), {\it or} towards a
 detector. According to the SLASH argument, a decrease of
 the photon detection rate in the first detector can be expected if
 the nonlinear device is inserted into the second beam pass.

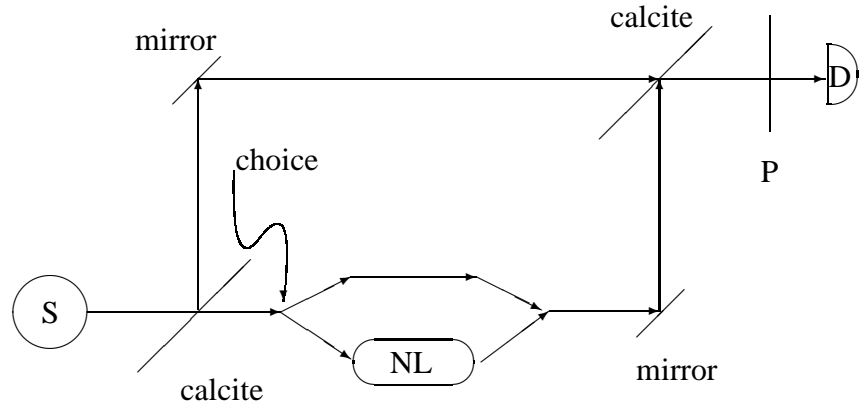
\begin{figure}
\begin{center}
\unitlength 1.00mm
\linethickness{0.4pt}
\begin{picture}(112.33,49.00)
\put(5.00,9.67){\circle{10.00}}
\put(5.00,9.67){\makebox(0,0)[cc]{S}}
\put(78.00,32.67){\line(1,1){15.00}}
\put(27.67,43.67){\line(-1,-1){6.33}}
\put(89.33,12.67){\line(-1,-1){6.33}}
\put(53.33,3.00){\oval(16.00,6.00)[]}
\put(108.17,41.33){\oval(8.33,8.00)[r]}
\put(108.34,45.33){\line(0,-1){8.00}}
\put(100.67,49.00){\line(0,-1){15.33}}
\put(53.00,3.00){\makebox(0,0)[cc]{NL}}
\put(100.67,28.33){\makebox(0,0)[cc]{P}}
\put(110.00,41.00){\makebox(0,0)[cc]{D}}
\put(10.00,9.67){\vector(1,0){25.67}}
\put(35.67,9.67){\vector(3,-2){9.33}}
\put(62.33,3.00){\vector(4,3){9.00}}
\put(71.33,9.75){\vector(1,0){14.67}}
\put(86.00,9.75){\vector(0,1){30.92}}
\put(24.67,40.67){\vector(1,0){61.33}}
\put(24.67,10.00){\vector(0,1){30.67}}
\put(35.67,9.67){\vector(2,1){9.33}}
\put(45.00,14.33){\vector(1,0){16.67}}
\put(61.67,14.33){\vector(2,-1){8.67}}
\put(36.00,11.33){\vector(0,-1){0.2}}
\bezier{92}(32.67,19.33)(37.67,26.00)(36.00,11.33)
\bezier{80}(32.67,19.33)(29.00,14.67)(29.33,28.33)
\put(29.67,30.00){\makebox(0,0)[lc]{choice}}
\put(21.67,46.00){\makebox(0,0)[cc]{mirror}}
\put(88.33,2.00){\makebox(0,0)[cc]{mirror}}
\put(85.00,49.00){\makebox(0,0)[cc]{calcite}}
\put(22.33,-0.67){\makebox(0,0)[lc]{calcite}}
\put(16.67,1.67){\line(1,1){15.00}}
\put(86.00,40.67){\vector(1,0){22.00}}
\end{picture}
\end{center}
\caption{\label{1989-slash-f3}
Study of the polarization of a recombined beam of an originally
 unpolarized light source in a Mach-Zehnder type interferometer.}
\end{figure}

 In yet another experimental setup (see Fig. \ref{1989-slash-f3}), two calcite crystals are
 used to recombine the beam from an originally unpolarized source.
 Inbetween those anomalous refractors, one polarized beam pass can
 optionally be intersected by the a nonlinear device. (The beam is not
 dumped into the device but {\it passes} it. In this respect the device
 acts merely as ``environment'' or ``populated vacuum''.) The
 polarization direction of the recombined light is then analyzed.
 It could be expected that if the nonlinear device intersects with the
 beam, then amplification of the polarization direction corresponding
 to the polarization direction of
 this nonlinear device is detected. (Qualitatively the same argument
 also holds if the calcite crystals are replaced by two semitransparent
 mirrors.)

Very similar arguments apply if the EPR-source radiates {\em fermions} instead of bosons.
In this case one could think of the nonlinear device as a box
of fermions in states which become forbidden for the EPR-decay.
Signalling is then obtained by {\em attenuation}.

      In summary it can be said that it is the author's believe that
 the proposal most certainly will turn out to be incorrect or
 unrealizable.  However, it is his hope that it yields stimulus and new
 insight to some aspects of the ongoing debate on the foundations of
 quantum theory, in particular with respect to quantum amplification,
the spatial behavior of stimulated emission,
 and the generation of noise.

 \section{Acknowledgements}
This work was supported in part by the
Erwin--Schr\"odinger--Gesellschaft f\"ur Mikrowissenschaften.
It was written long time ago in 1989,  distributed casually within the community,
and even made its way to a class at Yale University.
(There the task was to prove it is a hoax.)


\end{document}